\documentclass[twocolumn,showpacs,preprintnumbers,amsmath,amssymb]{revtex4} 
 
\usepackage{dcolumn} 
\usepackage{bm} 
\usepackage{helvet} 
 
\usepackage[dvips]{graphics} 
\usepackage[dvips]{graphicx} 
 
\usepackage{pst-all} 
\usepackage{pstcol} 
\usepackage{amsmath} 
\usepackage[english,activeacute]{babel} 
\usepackage{amssymb, latexsym} 
\usepackage{psfrag} 
\newpsobject{grilla}{psgrid}{subgriddiv=1,griddots=10,gridlabels=6pt} 
  
\begin{document}  
   
  \title{Atom chips with two-dimensional electron gases: theory of near surface trapping and\\ 
  ultracold-atom microscopy of quantum electronic systems}
  \author{G. Sinuco-Le\'on} 
  \author{B. Kaczmarek} 
  \author{P. Kr\"uger} 
  \author{T.M. Fromhold}  
  \affiliation {Midlands Ultracold Atom Research Centre, School of Physics and Astronomy, University of Nottingham, Nottingham, NG7 2RD, UK}
  \date{\today} 
     \begin{abstract}
We show that current in a two-dimensional electron gas (2DEG) can trap ultracold atoms $<1~\mu$m away with orders of magnitude less spatial noise than a metal trapping wire. This enables the creation of hybrid systems, which integrate ultracold atoms with quantum electronic devices to give extreme sensitivity and control: for example, activating a single quantized conductance channel in the 2DEG can split a Bose-Einstein condensate (BEC) for atom interferometry. In turn, the BEC offers unique structural and functional imaging of quantum devices and transport in heterostructures and graphene.  
\end{abstract} 
   
   
\maketitle 
Atom chips create microscopic potential landscapes for Bose-Einstein condensates (BECs) and degenerate Fermi gases \cite{chiprev,hinds,Thywissen}. High sensitivity of ultracold atoms to changes in the trap potential has led to chip-based atom interferometry \cite{Schumm2005}, field sensors \cite{brenton,Kurn} and BEC microscopes \cite{JAP65_361,wildermuth,brenton,Science319_1226}, which map current flow in classical metal conductors with $\mu$m-scale resolution. Atom chips have also been made using quantum coherent superconducting wires \cite{super1}. Semiconductor heterostructures \cite{pepper}, or graphene \cite{graphene}, containing a two-dimensional electron gas (2DEG) form another major class of quantum electronic devices. Such systems exhibit rich fundamental physics -- from the Quantum Hall Effects to single-photon sources and qubit manipulation -- and have technological applications in resistance standards and high-mobility transistors for mobile communication \cite{pepper}. Despite their transformative role in quantum electronics, there has been little discussion of 2DEGs in the context of atom chips. Hybrid matter-wave devices with controlled quantum coupling between ultracold atoms and 2DEG-based circuits offer many possibilities \cite{OurNJP}, but depend critically on achieving sub-micron trapping distances.
\begin{figure}[!htb] 
    \centering 
    \includegraphics[width=8.5cm]{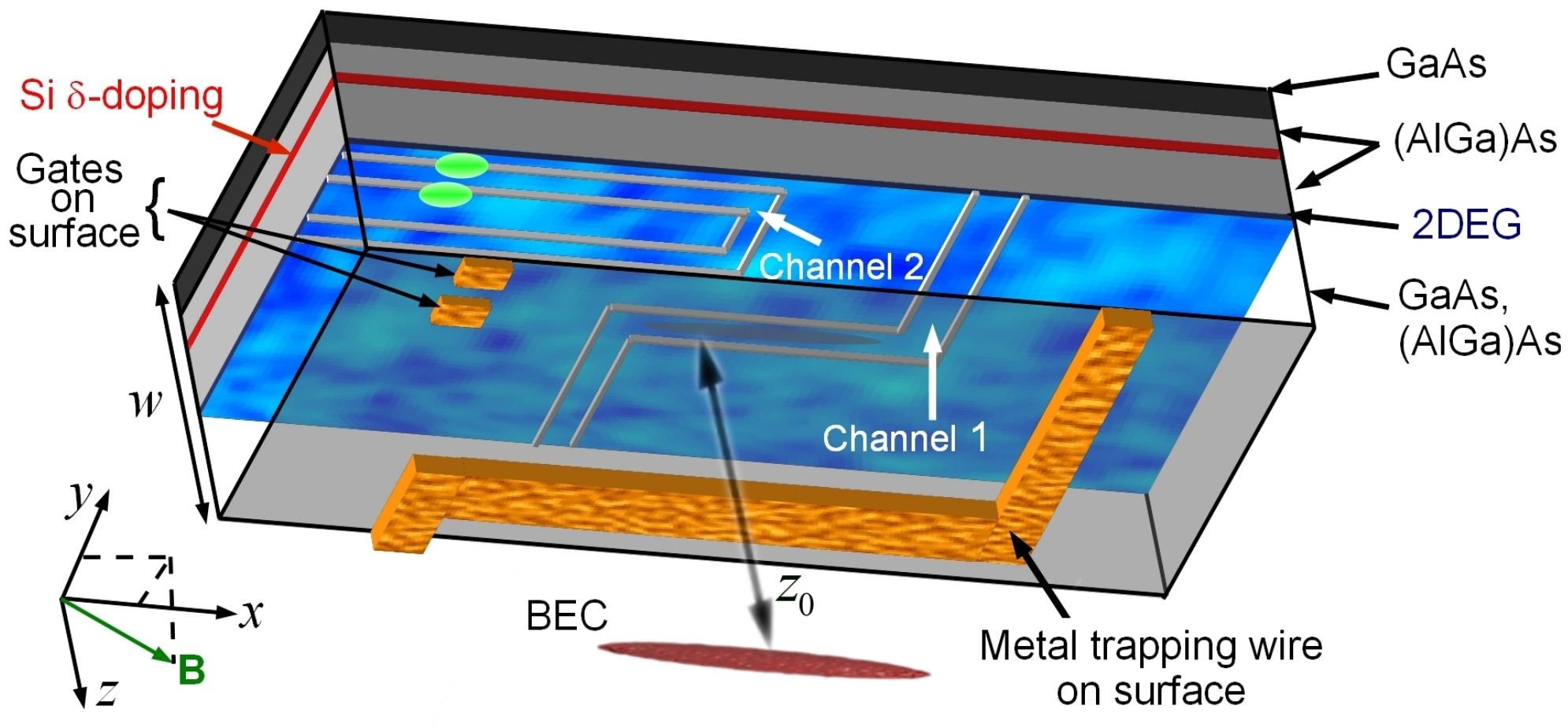}
    \caption{(Color) Schematic of a heterojunction-based atom chip comprising GaAs and (AlGa)As layers (labelled) in the $x-y$ plane (axes inset). Ionized Si donors (red layer) supply electrons to the 2DEG (blue), which lies in the $z=0$ plane. GaAs and (AlGa)As layers below the 2DEG are shown semitransparent for clarity. Insulating regions (gray) in the 2DEG enclose distinct conducting channels, 1 and 2, which are typically a few $\mu$m or nm wide respectively. A BEC (red) is confined in a harmonic trap, centered in the $z=z_0$ plane, produced by applied field, $\textbf{B}$, (green arrow), and current through either the metal surface wire (yellow Z shape) or 2DEG Channel 1. Rectangular yellow regions are metal surface gates, which can be biased negatively to produce a QPC (between light green depletion regions) in the lower arm of Channel 2.}    
    \label{geometry} 
  \end{figure} 
  
Sub-micron trapping has been achieved using evanescent light fields \cite{hammes} and may also be possible near ferromagnetic nanowires \cite{allwood}. But in all experiments to date on current-carrying atom chips, atom-surface trapping distances exceed $1~\mu$m. The common use of thick metal wires limits miniaturization of the potential landscape because of high Johnson noise \cite{noise1,noise2,noise3} and strong atom-surface Casimir-Polder (CP) attraction \cite{noise3}. Also, imperfections in the wires cause spatial fluctuations in the trapping potential, thus modulating the density profile of a BEC \cite{PRL92_076802,PhysRevA.76.063621,rough1,rough2,rough3}. Such fluctuations are undesirable from a trapping perspective and require oscillating rf currents to reduce them \cite{PhysRevA.76.063621,PRL98_263201}.
   
In this Letter, we show that 2DEG quantum electronic components fabricated in a GaAs/(AlGa)As heterojunction membrane \cite{pepper,Blick} overcome present limits on both the functionality and miniaturization of atom chips -- opening the way to integrating ultracold atoms with such components and broadening the scope of BEC microscopy \cite{wildermuth}. Inhomogeneities in the 2DEG current produce spatial fluctuations in the density profile of a BEC held nearby, which provide a direct \emph{non-invasive} measure of the current flow pattern, 2DEG potential landscape, and ionized donor distribution. This distribution is of fundamental importance for understanding and increasing 2DEG mobility \cite{Buks2,Coleridge,Grill,xray,Fleischmann}, but hard to determine directly in experiment. In contrast to metal wires \cite{PhysRevA.76.063621,EPJD32_171}, the variation of the rms amplitude of the density fluctuations, $\Delta {n}^{rms}$, with distance, $z$, from the 2DEG can be controlled \emph{in situ} by changing the distribution of ionized donors. Using optical illumination to pattern this distribution periodically makes $\Delta n^{rms}$ decrease exponentially with increasing $z$ so that it is \emph{3 orders of magnitude smaller} than for a metal wire $\approx 1~\mu$m from the chip. This, combined with low Johnson noise and weak CP attraction to membranes, makes 2DEGs -- in heterojunctions or in graphene -- ideal for producing the smooth stable near-surface traps required for creating hybrid cold-atom/quantum electronic systems. To illustrate the potential of such systems, we show that the BEC is so sensitive to the quantum conductor that it can detect the opening or closure of a \emph{single} quantized conductance channel in a quantum point contact (QPC) \cite{pepper,Fleischmann} -- enabling functional imaging of quantum electron transport and devices over hundreds of $\mu$m. Opening and closing the channel splits and remerges the BEC, demonstrating that delicate quantum electronic transport processes offer robust control of atomic matter waves.   

We consider an atom chip built on a GaAs/(AlGa)As heterojunction, which traps a BEC near a 2DEG (Fig.~\ref{geometry}). In the first part of the paper, the trapping potential is produced by currrent, $I_w$, through a Z-shaped metal surface wire (yellow in Fig.~\ref{geometry}), combined with a uniform magnetic field $\textbf{B}=(B_x, B_y, B_z)$, which positions the trap center at distance $z_0$ from conducting Channel 1 in the 2DEG \cite{wildermuth}. The 2DEG is formed by electrons from ionized donors in a Si $\delta$-doping layer (red in Fig.~\ref{geometry}) \cite{Buks2,Grill}, which migrate into the GaAs and populate the ground state of an almost triangular potential well formed at the GaAs/(AlGa)As interface \cite{pepper,Grill}. This confines the electrons in a narrow ($\approx$ 15 nm thick) sheet (blue in Fig.~\ref{geometry}). Insulating regions in the 2DEG (gray in Fig.~\ref{geometry}), made by implanting Ga ions \cite{ion_beam3}, enclose two distinct conduction channels, labelled 1 and 2 in Fig.~\ref{geometry}.   

Atom chips are usually built on a bulk substrate, which generates strong CP attraction, thus preventing atoms from being trapped closer than a few $\mu$m from the surface \cite{noise3}. Recently, suspended trapping wires were used to reduce the CP potential \cite{multilayer_chip}. We consider a heterojunction membrane of width $w=$ 130 nm, as in recent experiments \cite{Blick}, and calculate the CP potential energy $V_{CP}(z)$ [$\approx~-wC_5/z^5~(z~\gg w)$, where $C_5= 2 \times 10^{-54}$ Jm$^4$] via Eqs.~(25-29) of \cite{eberlein1}. As shown below, $V_{CP}$ is weak enough to allow submicron trapping.

We take typical heterojunction parameters with $n_d=3.3 \times 10^{15}$ $\text{m}^{-2}$ ionized donors at distance $d = 50$ nm from the 2DEG (Fig.~\ref{geometry}) \cite{OurNJP,PRB47_2233}. The 2DEG is $65$ nm below the surface and of mean electron density equal to $n_d$. Since the 2DEG is so thin, surface fluctuations are negligible: a major advantage over metal wires for near-surface trapping. We take the heterojunction temperature to be $4.2$ K (as in superconducting atom chips \cite{super1}) to ensure high 2DEG conductivity, $\sigma = 7.2 \times 10^{-2} \Omega ^{-1}$, and that inhomogeneity in the current originates only from non-uniformity of the ionized donors \cite{Buks2,Grill}.

These donors create a spatially-varying attractive potential, which is partially screened by the 2DEG \cite{pepper}. In the Thomas-Fermi screening model \cite{PRB47_2233}, the potential energy of a 2DEG electron at position $\textbf{r} =  (x,y)$ is    
  \begin{equation} 
    \Phi(\textbf{r}) = \frac{e^2}{4 \pi \epsilon \epsilon_0} \int  e^{-kd} 
    \frac{p_d(\textbf{k}) e^{i\textbf{k}\cdot\textbf{r}}}{k + k_s} d^2\textbf{k}, 
    \label{screened_potential} 
  \end{equation} 
where $\textbf{k} = (k_x,k_y)$, $k=\left|\textbf{k}\right|$, $p_d(\textbf{k})$ is the 2D Fourier transform of the spatial ionized donor density, $p_d(\textbf{r})=n_d+\Delta p_d(\textbf{r})$, $\epsilon$ is the relative permittivity of GaAs, and the screening wave vector, $k_s = e^2m^*/(2 \epsilon \epsilon_0 \pi \hbar^2)$, depends on the electron charge, $-e$, and effective mass, $m^*$ \cite{PRB47_2233}. 

\begin{figure}[!t] 
    \centering 
    \includegraphics[width=7.5cm]{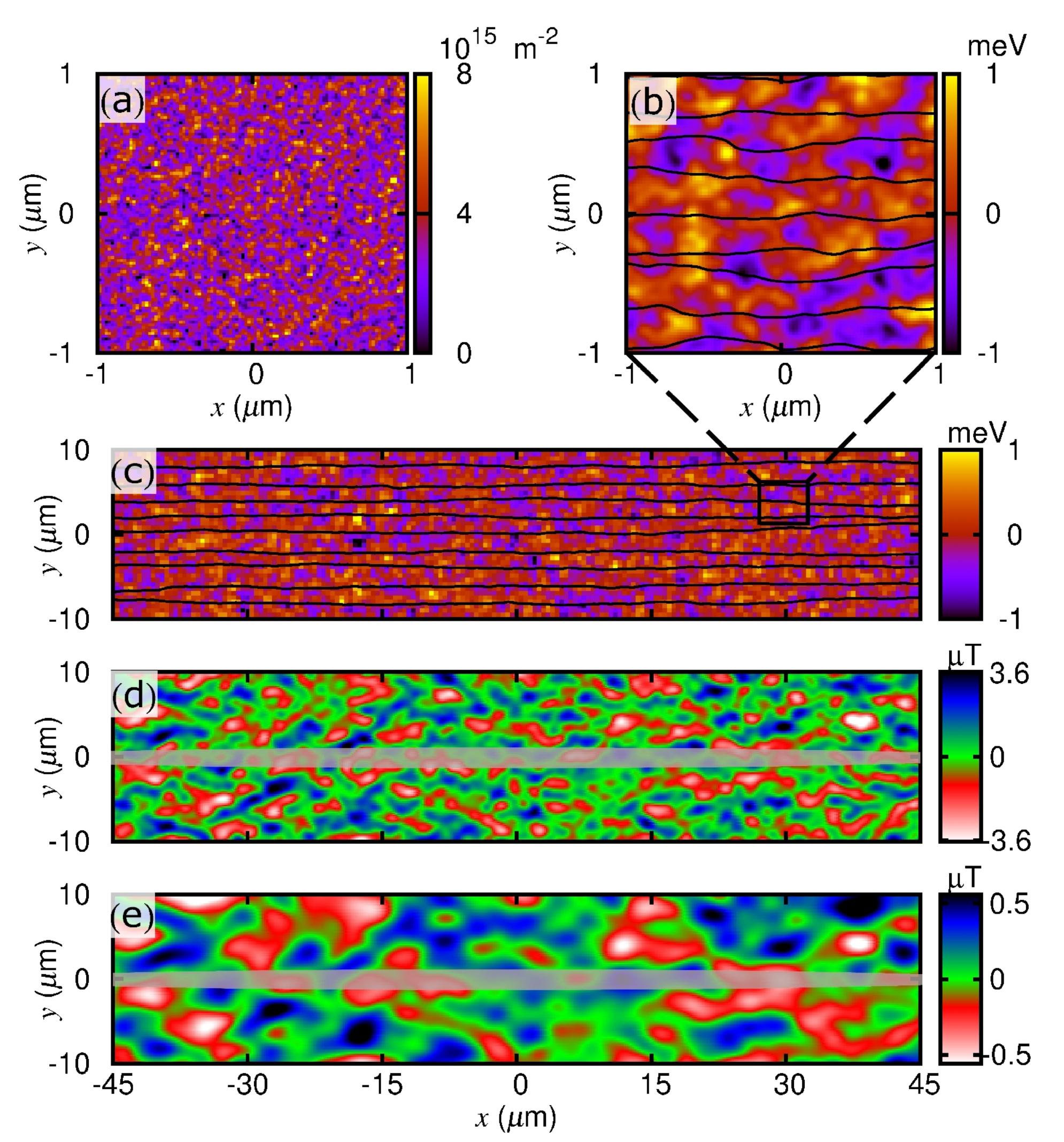}
    \caption{(Color) (a) Ionized donor density, $p_d(\textbf{r})$, where $\textbf{r}=(x,y)$. (b,c) Potential energy, $\Phi(\textbf{r})$, in the 2DEG. Region inside the rectangle in (c) is enlarged in (b). (d,e) Magnetic field component $B_x(\textbf{r},z_0)$ at distance $z_0=$ (d) $1~\mu$m, (e) $3~\mu$m from the 2DEG. Black curves in (b,c) are current streamlines. Gray shapes in (d,e): location of BEC \emph{A} in the $z=z_0$ plane.}   
    \label{two_deg_properties_fig_and_B_x_at} 
  \end{figure}

We now consider the electrostatic properties and current profile of Channel 1, whose width is henceforth taken to be $\gg z_0$. Figure~\ref{two_deg_properties_fig_and_B_x_at} shows (a) a typical uncorrelated ionized donor distribution, $p_d(\textbf{r})$, and (b,c) the screened potential energy in the 2DEG, $\Phi(\textbf{r})$ \cite{Coleridge}. 

When an applied voltage creates an electric field, $\textbf{E}$, along Channel 1, the local current density is $\textbf{j}(\textbf{r})= \sigma \textbf{E} + \Delta \textbf{j}(\textbf{r})$ \cite{streamline}, where the inhomogenous component $\Delta \textbf{j}(\textbf{r})= \sigma \nabla \Phi(\textbf{r})/e$ originates from fluctuations in $p_d(\textbf{r})$ and $\Phi(\textbf{r})$.

The black curves in Fig.~\ref{two_deg_properties_fig_and_B_x_at}(b,c) show current streamlines calculated for $E = 2.5$ kVm$^{-1}$. Imhomogeneity of $\Phi(\textbf{r})$ makes these streamlines deviate from the $x-$direction. Consequently, the resulting magnetic field at position $\textbf{r}=(x,y)$ in the $z=z_0$ plane has a non-zero $x-$component given, from the Biot-Savart law \cite{JAP65_361,streamline}, by  
  \begin{equation} 
    B_x(\textbf{r},z_0)= \frac{\mu_0 e \sigma}{4 \epsilon \epsilon_0} \int \frac{k_y \ \Delta p_d(\textbf{k})}{k + k_s} e^{-k(d+z_0)} e^{i \textbf{k} \cdot \textbf{r}} d^2\textbf{k}, 
    \label{Biot-Savart-donor_dist} 
  \end{equation} 
where $\Delta p_d(\textbf{k})$ is the 2D Fourier transform of the donor density fluctuations, $\Delta p_d(\textbf{r})$. Equation (\ref{Biot-Savart-donor_dist}) reveals two key results. First, the fluctuations of $B_x(\textbf{r},z_0)$ depend directly on the donor distribution via $\Delta p_d(\textbf{k})$. Second, due to the $e^{-k(d+z_0)}$ term, the field fluctuations become weaker and smoother as $z_0$ increases. This can be seen by comparing the color maps of $B_x(\textbf{r},z_0)$ in Fig.~\ref{two_deg_properties_fig_and_B_x_at}(d,e). 

We now consider how the field fluctuations affect a BEC, henceforth called BEC \emph{A}, comprising $10^4$ $^{87}$Rb atoms in state $\left|F=2,m_F=2\right\rangle$, whose position in the $z=z_0$ plane is shown by the gray regions in Fig.~\ref{two_deg_properties_fig_and_B_x_at}(d,e). Since the BEC is strongly confined along $y$ and $z$, its density profile is sensitive only to fluctuations in $B_x(\textbf{r},z_0)$. Along the $x-$axis [$\textbf{r}=(x,0)$], the atom density fluctuations are $\Delta n(x) = -(m_F g_F \mu_B/2 \hbar \omega_{r}a_s)B_x(x,z_0)$ \cite{wildermuth}, where the g-factor $g_F=1/2$, $\mu_B$ is the Bohr magneton, $\omega_{r}$ is the trap frequency in the $y-z$ plane and $a_s$ is the s-wave scattering length. 

The variation of the oscillatory amplitudes of $B_x(x,z_0)$ with $z_0$ is crucial for understanding, and exploiting, the effect of 2DEG current on a BEC. To demonstrate this, we first consider the rms average, $B_{x}^{rms}(z_0)$, of $B_x(x,z_0)$ along $x$ at given $z_0$. If the spatial donor correlation function, $\left\langle \Delta p_d(\textbf{r}) \Delta p_d(\textbf{r}')\right\rangle=C(\textbf{r},\textbf{r}')=C(\left|x-x'\right|,y,y')$, is homogeneous along $x$ and $\rightarrow 0$ as $\left|x-x'\right|\rightarrow\infty$, the rms spatial average is equivalent to an ensemble average, denoted $\left\langle ...\right\rangle$, at any given $x$. Choosing $x=0$ gives  
\begin{eqnarray}
B_{x}^{rms}(z_0) = \left( \frac{\mu_0 e \sigma}{4 \epsilon \epsilon_0} \right) \left [\int \int \frac{k_{y} k'_y S(\textbf{k},\textbf{k}')}{(k+k_s)(k'+k_s)} \right. \nonumber \\ \left.
\times e^{-(k+k')(d+z_0)} d^2\textbf{k}d^2\textbf{k}' \right]^{1/2}.
\label{Biot-Savart-donor_dist_B}
\end{eqnarray}
Equation (\ref{Biot-Savart-donor_dist_B}) reveals that $B_{x}^{rms}(z_0)$ depends on, and can hence probe, the correlation function of the ionized donors, $S(\textbf{k},\textbf{k}') = \left\langle \Delta p_d(\textbf{k}) \Delta p_d(\textbf{k}') \right\rangle$, in \textbf{k}-space. 

\begin{figure}[!t] 
    \centering 
    \includegraphics[width=6.5cm]{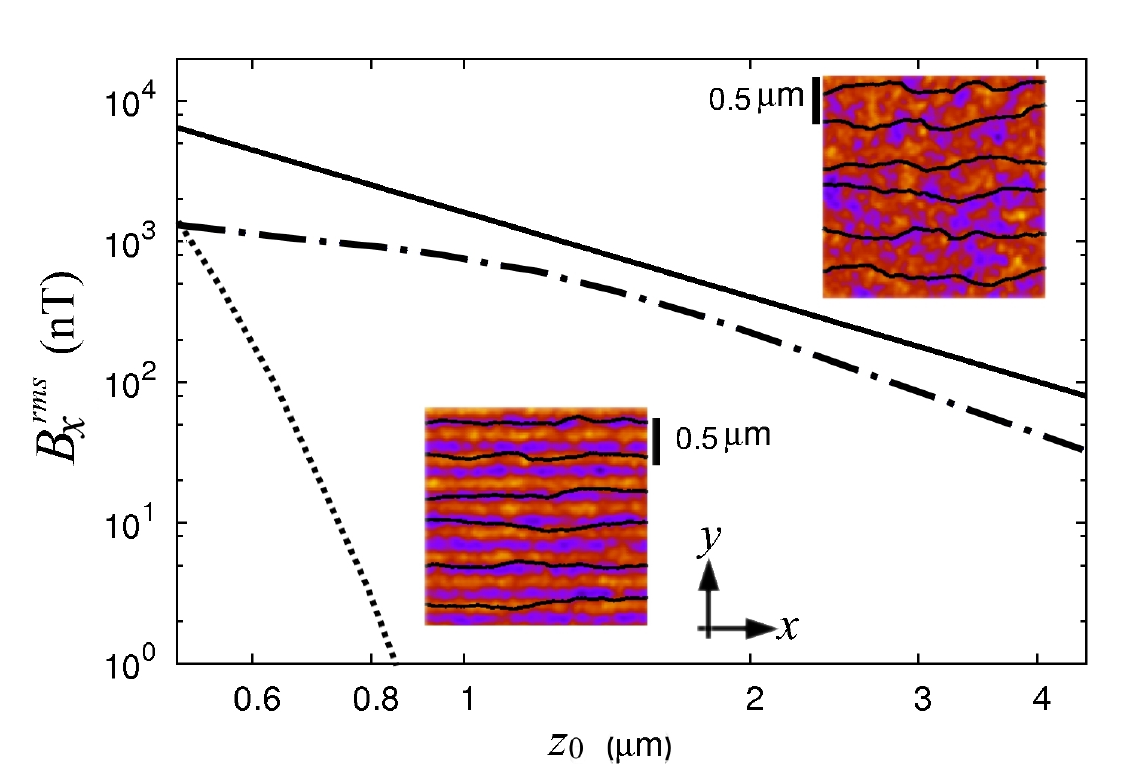} 
    \caption{(Color) $B_{x}^{rms}(z_0)$ above Channel 1 calculated for spatially random (solid curve) and periodically-modulated (dotted curve) ionized donor profiles in $x-y$ plane (right- and left-hand insets respectively: black curves are current streamlines at $E = 1.6$ kVm$^{-1}$ and bars show scale). Dot-dashed curve: $B_{x}^{rms}(z_0)$ calculated above a metal wire \cite{EPJD32_171}.}
      \label{B_x_decay_BS}  
  \end{figure}

For a random donor distribution (Fig.~\ref{B_x_decay_BS} right inset) \cite{Coleridge}, $S(\textbf{k},\textbf{k}') \propto \delta(\textbf{k}+\textbf{k}')$. Using this in Eq. (\ref{Biot-Savart-donor_dist_B}) gives $B_{x}^{rms}(z_0) \sim 1/z_0^2$ (solid curve in Fig.~\ref{B_x_decay_BS}), meaning that for $z_0 \gtrsim 1.5~\mu$m, field fluctuations decay more slowly above a 2DEG than above a metal wire (dot-dashed curve in Fig.~\ref{B_x_decay_BS}, calculated for a wire of width $\gg z_0$ using formulae and parameters for surface and edge roughness from \cite{EPJD32_171}). As a result, the BEC responds to field fluctuations over a range of $z_0$, with an upper limit $\sim 4~\mu$m where $\Delta n(x)$ falls too low to detect ($\lesssim 10\%$ of the mean atom density \cite{Science319_1226}), and a lower limit $\sim 700~$nm determined by $V_{CP}(z)$ (see below). The form of $B_{x}^{rms}(z_0)$ gives information about donor correlations in this range.  

Both $p_d(x,y)$ and $\Phi(x,y)$ can be determined directly from measurements of $B_x(x,y)$ made by scanning the BEC over the 2DEG at \emph{fixed} $z_0$, which sets the resolution \cite{JAP65_361,wildermuth,brenton,Science319_1226}. As shown in \cite{wildermuth}, a Fourier transform of $B_x(x,y)$ gives $\Delta j_y(x,y)$: $\Delta j_x(x,y)$ follows from current continuity, thence $\Phi(x,y)$ from integrating $\nabla \Phi(x,y) \propto \Delta \textbf{j}(x,y) $. Finally, deconvolving the relation between $\Phi(x,y)$ and $p_d(\textbf{k},\textbf{k}')$ stated in \cite{PRB47_2233}, using the method in \cite{wildermuth}, gives $p_d(\textbf{k},\textbf{k}')$, $S(\textbf{k},\textbf{k}')$, and $p_d(x,y)$. For $z_0 \lesssim 1~\mu$m, the resolution is sufficient to image, for example, electron redistribution during metal-insulator transitions \cite{Ilani,Grill} and, in particular, $\mu$m-scale ionized donor correlations in high mobility 2DEGs \cite{Mico,High_mob,siegert}. Donor statistics are of fundamental importance to electron transport and mobility \cite{Buks2,Coleridge,Grill,xray,Fleischmann}, but usually inferred from theoretical models due to a lack of non-invasive experimental probes \cite{Grill,Fleischmann}. Scanning probe methods can image 2DEGs on the nm scale over a few $\mu$m$^2$. In contrast, BEC microscopy offers non-invasive, fast (single-shot) imaging of $p_d(x,y)$ and $\Phi(x,y)$ over large (100s of $\mu$m$^2$) areas of the 2DEG. So it is ideal for imaging, \emph{in situ}, long-range donor correlations, which are the key to enhancing mobility, and their response to illumination or thermal cycling.

The ionized donor profile, and hence $B_{x}^{rms}(z_0)$, can be permanently changed by transient optical illumination \cite{xray}. Since the ionized donor density increases with the local light intensity, two laser beams, of wavelength $\lambda$, counter-propagating along $y$, spatially modulate the donor profile with a period $\lambda_y=\lambda/2$ (Fig.~\ref{B_x_decay_BS} left inset). In this case, $S(\textbf{k},\textbf{k}') \propto [\delta(k_y+k_0)\delta(k_y'-k_0) + \delta(k_y-k_0)\delta(k_y'+k_0)]$, where $k_0 = 2 \pi / \lambda_y$. Using this in Eq. (\ref{Biot-Savart-donor_dist_B}) gives $B_{x}^{rms}(z_0) \sim \exp(-z_0/ \lambda_y)$, shown for $\lambda =$ 660 nm by the dotted curve in Fig.~\ref{B_x_decay_BS}. This rapid decay occurs because, as in a magnetic mirror \cite{hinds}, the current streamlines align with the striped donor pattern (Fig.~\ref{B_x_decay_BS} left inset), which reduces their meander and, hence, $B_{x}^{rms}(z_0)$. 

The ability to tailor the potential landscape of the 2DEG, and the resulting field fluctuations, is a unique feature of heterojunctions, which can be exploited for trapping, manipulating, and imaging with, ultracold Bose gases. Crucially, exponential decay makes the $B_{x}^{rms}(z_0)$ curve for the periodically-modulated donor distribution (dotted in Fig.~\ref{B_x_decay_BS}) fall rapidly below that for a metal wire (dot-dashed curve in Fig.~\ref{B_x_decay_BS}). At $z_0\gtrsim 0.8~\mu$m, the field fluctuations above the 2DEG are more than \emph{3 orders of magnitude smaller} than for the metal wire.  

\begin{figure}
    \centering 
   \includegraphics[width=8.5cm]{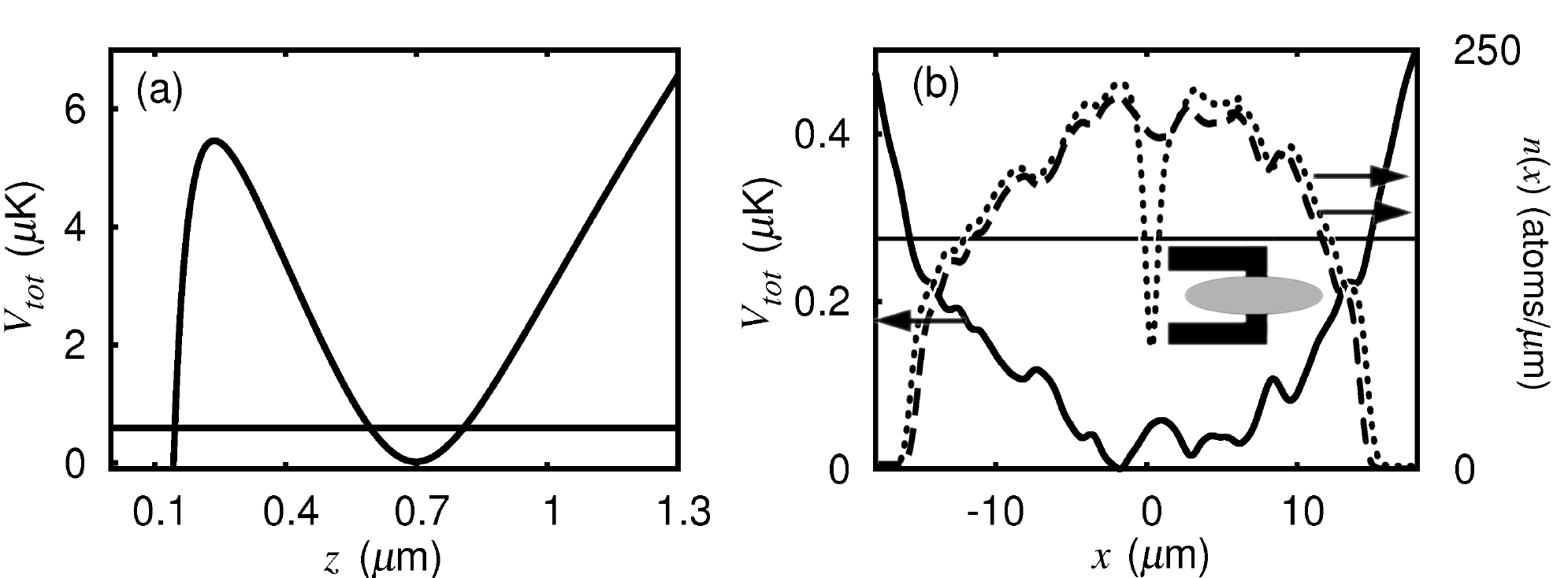} 
   \caption{(a,b): $V_{tot}$ (solid curve) for a $^{87}$Rb atom versus (a) $z$, (b) $x$ at $z_0=0.7~\mu$m above the center of Channel 1 (Fig.~\ref{geometry}). Horizontal lines in (a,b): chemical potential of BEC \emph{B} in the trap. In (b), dashed (dotted) curve shows for $N$ = 0 (1) the density of BEC \emph{B}, $n(x)$, along the trap center, which is displaced so that the BEC (schematic gray shape in inset) lies above the central arm (black in inset) of Channel 2.}
      \label{QPC_BEC}  
  \end{figure}

Consequently, 2DEGs have great potential for creating near-surface microtraps. To demonstrate this, we consider the trap produced by a 355$~\mu$A current through 2DEG Channel 1 (of width $3~\mu$m and central arm length $60~\mu$m) \emph{only}, setting $I_w=0$. The solid curve in Fig.~\ref{QPC_BEC}(a) shows the total potential energy $V_{tot}(z)=V_m(z)+V_{CP}(z)$ calculated for an $^{87}$Rb atom, where $V_{m}(z)$ originates from the magnetic field produced by Channel 1 and an applied field $\textbf{B}=(40, 536, 0)$ mG. Since the CP attraction is weak, the trap is deep enough to confine BEC \emph{B}, comprising $500$ $^{87}$Rb atoms with $\left|F=2,m_F=2 \right\rangle$, whose chemical potential [horizontal line in Fig.~\ref{QPC_BEC}(a)] is far below the top of the left-hand barrier. Even though the trap center is only $0.7~\mu$m from the 2DEG [Fig.~\ref{QPC_BEC}(a)], $V_{tot}$ varies fairly smoothly with $x$ [solid curve in Fig.~\ref{QPC_BEC}(b)] because $B_{x}^{rms}(z_0)$ is only $\approx 20~$nT (Fig.~\ref{B_x_decay_BS}). Even at 4.2 K, the resistance of Channel 1 is $\approx$ 10 times higher than a similarly-sized gold conductor at room temperature \cite{Folman_sub}. Consequently, 2DEGs offer lower Johnson noise and spin-flip loss rates than metal wires. 

Near-surface trapping makes the BEC highly sensitive to magnetic field variations arising from the \emph{geometry} of the conduction channels, such as local narrowing. As an example, suppose that \textbf{B} is adjusted to hold BEC \emph{B} across the middle arm of U-shaped Channel 2 [Figs.\ \ref{geometry} and \ref{QPC_BEC}(b) inset]. Channel 2 is narrow enough (20 nm) for the electrons to populate a small number, $N$, of discrete energy levels, $E_l$ ($l= 0,1,2,..., N$), corresponding to motion across it. Along the channel, whose quantized conductance is $G=2Ne^2/h$ \cite{pepper,xray}, electrons occupy one-dimensional plane wave states up to the Fermi level, $E_F$. Applying a negative voltage, $-\left| V_g\right|$, to metal surface gates (yellow rectangles in Fig.~\ref{geometry}), which are on either side of the lower arm of Channel 2 and sufficiently far from the BEC to have negligible electrostatic effect on it, locally depletes the 2DEG. This narrows the conduction channel, forming a QPC and raising $E_l$. As $V_g$ increases, the energy levels successively exceed $E_F$ and depopulate \cite{pepper,xray,Fleischmann}. Depopulation of each level decreases $N$ and, hence, the current through Channel 2 by $\Delta I=2e^2V/h$, where $V$ is the voltage dropped across the QPC. 
  
The dashed curve in Fig.~\ref{QPC_BEC}(b) shows the density profile, $n(x)$, of BEC \emph{B} when the QPC in Channel 2 is fully depleted ($N=0$). Since the QPC carries no current in this case, $n(x)$ is the unperturbed ground state of the trapping potential shown by the solid curves in Fig.~\ref{QPC_BEC}(a,b). Opening \emph{a single} conduction channel in the QPC ($N=1$) changes the trap profile sufficiently to almost completely split the BEC [dotted curve in Fig.~\ref{QPC_BEC}(b)]. Consequently, the BEC can detect quantized changes in the QPC's conductance, which, conversely, can manipulate the BEC, for example splitting and recombining it in atom interferometry. More complex shaping of the atom cloud is also possible: an array of QPCs could imprint and control a wide range of sub-$\mu$m patterns in the BEC.

In summary, quantum electron transport in heterojunctions can create smooth, low-noise, magnetic traps, which provide the sub-$\mu$m control required to integrate ultracold atoms with quantum electronic systems. By tailoring the donor distribution, magnetic field fluctuations can be suppressed exponentially. Quantum transport processes in a 2DEG can imprint strong density modulations in a BEC, which, conversely, provide a non-invasive probe of those processes. The ability to measure the potential landscape and mobility of a 2DEG \emph{independently} may yield new insights for understanding how the two relate and, hence, for increasing 2DEG mobility \cite{Mico,High_mob}. Suspended graphene membranes, which combine low CP attraction with high room-temperature mobility \cite{graphene}, could be the ultimate material for sub-$\mu$m atom trapping and BEC microscopy of quantum transport.

\begin{thebibliography}{41}
\bibitem[{str()}]{streamline}
\bibinfo{note}{Assuming \textbf{E} is large enough for $\Phi(\textbf{r})$ to
  weakly perturb the current streamlines. Atom density fluctuations then result
  only from $\Delta j_y(\textbf{r})$ and are thus independent of \textbf{E}.}
\end{thebibliography}

\end{document}